\documentclass[aps,prl,twocolumn,showpacs]{revtex4}

\def\beq{\begin{equation}}
\def\eeq{\end{equation}}
\def\bea{\begin{eqnarray}}
\def\eea{\end{eqnarray}}
\def\nn{\nonumber}
\def\sss{\scriptscriptstyle}
\def\bd{B_d^0}

\def\bs{B_s^0}
\def\bsbar{{\overline{B_s^0}}}
\def\ks{K_{\sss S}}

\def\roughly#1{\mathrel{\raise.3ex\hbox
{$#1$\kern-.75em\lower1ex\hbox{$\sim$}}}}
\def\lsim{\roughly<}

\def\pewp{P'_{\sss EW}}
\def\pewcp{P_{\sss EW}^{\sss \prime C}}
\def\btopik{B \to \pi K}
\def\btod{{\bar b} \to {\bar d}}
\def\btos{{\bar b} \to {\bar s}}
\def\ApNPqph{{\cal A}^{\prime,q} e^{i \Phi'_q}}
\def\ApNPuph{{\cal A}^{\prime,u} e^{i \Phi'_u}}
\def\ApNPdph{{\cal A}^{\prime,d} e^{i \Phi'_d}}
\def\ApNPCqph{{\cal A}^{\prime {\sss C}, q} e^{i \Phi_q^{\prime C}}}
\def\ApNPCuph{{\cal A}^{\prime {\sss C}, u} e^{i \Phi_u^{\prime C}}}
\def\ApNPCdph{{\cal A}^{\prime {\sss C}, d} e^{i \Phi_d^{\prime C}}}
\def\ApNPcomb{{\cal A}^{\prime, comb} e^{i \Phi'}}

\def\bra#1{\left\langle #1\right|}
\def\ket#1{\left| #1\right\rangle}
\def\ANPq{{\cal A}^q}

\begin{document}

\preprint{UdeM-GPP-TH-04-128}
\title{The $\btopik$ Puzzle and New Physics}
\author{Seungwon Baek}
\author{Philippe Hamel}
\author{David London}
\affiliation{ Laboratoire Ren\'e J.-A. L\'evesque, 
Universit\'e de Montr\'eal, C.P. 6128, succ.~centre-ville, Montr\'eal, QC,
Canada H3C 3J7} 
\author{Alakabha Datta}
\affiliation{Department of Physics, University of Toronto,
60 St.\ George Street, Toronto, ON, Canada M5S 1A7}
\author{Denis A. Suprun}
\affiliation{Physics Department, Brookhaven National Laboratory,
Upton, NY, USA, 11973}
\date{\today}
\begin{abstract}
The present $\btopik$ data is studied in the context of the standard
model (SM) and with new physics (NP). We confirm that the SM has
difficulties explaining the $\btopik$ measurements. By adopting an
effective-lagrangian parametrization of NP effects, we are able to
rule out several classes of NP. Our model-independent analysis shows
that the $\btopik$ data can be accommodated by NP in the electroweak
penguin sector.
\end{abstract}

\maketitle

The $B$-factories BaBar and Belle have measured (most of) the
branching ratios and CP asymmetries for the various $B\to\pi\pi$ and
$\btopik$ decays, and these can be used to search for physics beyond
the Standard Model (SM). By using flavor SU(3) symmetry to relate
these processes \cite{BFRS,CGRS,MY,WZ,CCS,CL,HM}, several analyses
were able to constrain the SM parameters, and to look for signs of New
Physics (NP). The advantage of this approach is that one takes into
account a large number of processes. The disadvantage is that one has
to deal with unknown effects related to the breaking of SU(3)
symmetry. Also, $B\to\pi\pi$ decays involve the quark-level processes
$\btod q {\bar q}$ ($q=u,d$), while $\btopik$ receives contributions
from $\btos q {\bar q}$. If there is NP, it could affect $\btod$ and
$\btos$ processes differently.

For this reason, there are advantages to considering $\btopik$ decays
alone. As we will see, these processes contain enough information to
constrain the SM parameters. Within the diagrammatic approach
\cite{GHLR}, the amplitudes for the four $\btopik$ decays can be
written in terms of seven diagrams: the color-favored and
color-suppressed tree amplitudes $T'$ and $C'$, the gluonic penguin
amplitudes $P'$ and $P'_{uc}$, the color-favored and color-suppressed
electroweak penguin amplitudes $\pewp$ and $\pewcp$, and the
annihilation amplitude $A'$. (The primes on the amplitudes indicate
$\btos$ transitions.)

In Ref.~\cite{GHLR}, the relative sizes of the amplitudes were roughly
estimated as
\bea
1 & : & |P'|, \nn\\
{\cal O}({\bar\lambda}) & : & |T'|,~|\pewp|, \nn\\
{\cal O}({\bar\lambda}^2) & : & |C'|,~|P'_{uc}|,~|\pewcp|, \nn\\
{\cal O}({\bar\lambda}^3) & : & |A'|,
\eea
where ${\bar\lambda} \sim 0.2$. These estimates are expected to hold
approximately in the SM. Thus, to $O({\bar\lambda})$, we can ignore
all diagrams but $P'$, $T'$ and $\pewp$ in our $\btopik$
amplitudes. We will perform a fit of the present $\btopik$ data -- the
goodness or badness of the fit should not be much affected by the
inclusion of the smaller amplitudes.

The four amplitudes can then be written as
\bea
\label{SMamps}
A(B^+ \to \pi^+K^0) \equiv A^{+0} & = & -P' ~, \nn\\
\sqrt{2} A(B^+ \to \pi^0K^+) \equiv \sqrt{2} A^{0+} & = & -T'
e^{i\gamma} +P' - \pewp , \nn\\
A(B^0\to \pi^-K^+) \equiv A^{-+} & = & -T' e^{i\gamma} +P' ~, \nn\\
\sqrt{2} A(B^0\to \pi^0K^0) \equiv \sqrt{2}A^{00} & = & -P' - \pewp ~,
\eea
where we have explicitly written the dependence on the weak phase
(including the minus sign from $V_{tb}^* V_{ts}$ [$P'$]), while the
amplitudes contain strong phases. (The phase information in the
Cabibbo-Kobayashi-Maskawa (CKM) quark mixing matrix is conventionally
parametrized in terms of the unitarity triangle, in which the interior
(CP-violating) angles are known as $\alpha$, $\beta$ and $\gamma$
\cite{pdg}.)

We have one additional piece of information: within the SM, to a good
approximation, the diagram $\pewp$ can be related to $T'$ using flavor
SU(3) \cite{EWPs}:
\beq
\pewp \simeq {3\over 4} \left[ {c_9 + c_{10} \over c_1 + c_2} + {c_9 -
c_{10} \over c_1 - c_2} \right] R \, T' ~.
\label{pewprel}
\eeq
Here, the $c_i$ are Wilson coefficients \cite{BuraseffH} and 
\beq
R \equiv \left\vert {V_{tb}^* V_{ts} \over V_{ub}^* V_{us}}
\right\vert = {1 \over \lambda^2} {\sin(\beta +\gamma) \over
\sin\beta}~.
\eeq

With the above relation, the $\btopik$ observables contain five
theoretical parameters: $|P'|$, $|T'|$, $\beta$, $\gamma$, and one
relative strong phase, $\delta$. The phase $\beta$ can be taken from
the measurement of $\sin 2\beta$ in $\bd(t) \to J/\psi\ks$: $\sin
2\beta = 0.726 \pm 0.037$ \cite{sin2beta}, leaving four theoretical
unknowns. However, there are a total of nine $\btopik$ measurements:
four CP-averaged branching ratios and five CP asymmetries
(Table~\ref{bpiktable}, \cite{piKrefs}). Within the parametrization of
Eq.~(\ref{SMamps}), three of these are independent of the theoretical
parameters: the direct CP asymmetries in $B^+ \to \pi^+K^0$ and
$B^0\to \pi^0K^0$ are predicted to vanish, and the indirect CP
asymmetry in $B^0\to \pi^0K^0$ measures $\sin 2\beta$. The remaining
six observables are functions of the four theoretical parameters, so
we can perform a fit to obtain these quantities.

\begin{table}
\caption{Branching ratios, direct CP asymmetries $A_{dir}$, and
mixing-induced CP asymmetry $A_{indir}$ (if applicable) for the four
$\btopik$ decay modes.}
\center
\begin{tabular}{ c c c c }
Mode & $BR(10^{-6})$ & $A_{dir}$ & $A_{indir}$ \\ \hline
$B^+ \to \pi^+ K^0$ & $24.1 \pm 1.3$ & $-0.020 \pm 0.034$ & \\
$B^+ \to \pi^0 K^+$ & $12.1 \pm 0.8$ & ~~$0.04 \pm 0.04$ & \\
$\bd \to \pi^- K^+$ & $18.2 \pm 0.8$ & $-0.109 \pm 0.019$ & \\
$\bd \to \pi^0 K^0$ & $11.5 \pm 1.0$ & $-0.09 \pm 0.14$ & $0.34 \pm
0.28$ \\
\hline\hline
\end{tabular}
\label{bpiktable}
\end{table}

Using the parametrization of Eq.~(\ref{SMamps}) for the $\btopik$
amplitudes, we find that $\chi^2_{min}/d.o.f. = 15.6/5$ (0.8\%),
indicating an extremely poor fit. (The number in parentheses indicates
the quality of the fit, and depends on $\chi^2_{min}$ and $d.o.f.$
individually. 50\% is a good fit; fits which are substantially less
than 50\% are poor.) This is not a new result -- other analyses have
made a similar observation \cite{BFRS,MY,WZ,CCS,CL,HM}.  It shows that
present data is inconsistent with the naive implementation of the SM.
Our parametrization is therefore incomplete. There are two ways to
make modifications. Either we work within the SM, or we add new
physics. We address these possibilities in turn.

We begin with the SM, but abandon the relation between $\pewp$ and
$T'$ [Eq.~(\ref{pewprel})]. We now have six theoretical parameters:
$|P'|$, $|T'|$, $|\pewp|$, $\gamma$, and two relative strong
phases. In this case, the fit is good: $\chi^2_{min}/d.o.f. = 2.7/3$
(44\%). The fit also gives a central value of $\gamma =
59^\circ$. This is consistent with the value of $\gamma$ obtained via
a fit to independent measurements: $\gamma = {62^{+10}_{-12}}^\circ$
\cite{CKMfitter}. (Because these errors are not gaussian, we do not
include this information in our fit at this stage.) On the other hand,
the fit also gives $|\pewp/T'| = 1.55$, which is far from its SM value
of $0.65 \pm 0.15$ \cite{EWPs}. Thus, while it is possible to explain
the present $\btopik$ data by treating $\pewp$ and $T'$ independently,
it is difficult to understand how $|\pewp/T'|$ could be $6\sigma$ away
from its SM value.

The second modification is to take into account the smaller
(neglected) amplitudes. Including the $O({\bar\lambda}^2)$ diagrams,
the $\btopik$ amplitudes take the form
\bea
A^{+0} & = & -P' + P'_{uc} e^{i\gamma} -\frac13 \pewcp ~, \nn\\
\sqrt{2} A^{0+} & = & -T' e^{i\gamma} -C' e^{i\gamma} +P' - P'_{uc}
e^{i\gamma} \nn\\
& & \hskip30truemm -~\pewp -\frac23 \pewcp ~, \\
A^{-+} & = & -T' e^{i\gamma} + P' -P'_{uc} e^{i\gamma} -\frac23 \pewcp
~, \nn\\
\sqrt{2}A^{00} & = & -C' e^{i\gamma} - P' +P'_{uc} e^{i\gamma} - \pewp
-\frac13 \pewcp ~. \nn
\eea
In this case, $\pewcp$ is not independent of the amplitudes $T'$ and
$C'$. We have \cite{EWPs}
\bea
\pewp & \!\!=\!\! & {3\over 4} {c_9 + c_{10} \over c_1 + c_2} R (T' +
C') \!+\!  {3\over 4} {c_9 - c_{10} \over c_1 - c_2} R (T' - C')
~, \nn\\
\pewcp & \!\!=\!\! & {3\over 4} {c_9 + c_{10} \over c_1 + c_2} R (T' +
C') \!-\!  {3\over 4} {c_9 - c_{10} \over c_1 - c_2} R (T' - C')
~.
\eea

With these relations, we now have eight theoretical parameters:
$|P'|$, $|P'_{uc}|$, $|T'|$, $|C'|$, $\gamma$, and three relative
strong phases. With nine pieces of experimental data, we can still
perform a fit, which is acceptable: $\chi^2_{min}/d.o.f. = 0.7/1$
(40\%). In addition, we find $\gamma = 64^\circ$, consistent with
independent measurements. However, the fit gives $|C'/T'| = 1.8$,
which is far larger than naive estimates. Other analyses have also
found the $C'$ must be very big to explain the $\btopik$ data
\cite{WZ,CCS,CL,HM}. In Ref.~\cite{CCS}, it is argued that final-state
interactions (FSI) can increase the size of $C'$. However, in that
case, the authors were attempting to explain $|C'/T'| \simeq 0.5$
(which comes from the joint fit to $\btopik$ and $B\to\pi\pi$ decays
\cite{CGRS}). Even with FSI, it is difficult to see how $C'$ can be
increased to about twice as large as $T'$.

It is therefore extremely difficult to explain the current $\btopik$
data within the SM alone. Instead, one must consider the effect of
new-physics operators. These can be included in $B$-physics analyses
in a model-independent way \cite{DLNP}. We suppose that there are NP
contributions to $\btos q {\bar q}$ transitions which are roughly the
same size as the SM $\btos$ penguin operators. The NP contributions
take the form ${\cal O}_{\sss NP}^{ij,q} \sim {\bar s} \Gamma_i b \,
{\bar q} \Gamma_j q$ ($q = u,d,s,c$), where the $\Gamma_{i,j}$
represent Lorentz structures, and colour indices are suppressed. There
are a total of 20 possible NP operators, each of which can in
principle have a different weak phase. The NP contributes to the decay
$B\to f$ through its matrix elements $\bra{f} {\cal O}_{\sss
NP}^{ij,q} \ket{B}$, which can be written as
\beq
\bra{f} {\cal O}_{\sss NP}^{ij,q} \ket{B} = A_k e^{i \phi_k^q} e^{i
\delta_k^q} ~,
\eeq
where $\phi_k^q$ and $\delta_k^q$ are the NP weak and strong phases
associated with the individual matrix elements. However, the key point
is that the NP strong phases are very small. The reasoning goes as
follows. Strong phases arise from rescattering. In the SM, the (large)
tree diagram (${\widetilde T}'$) $\btos c{\bar c}$ can rescatter into
the $c$-quark penguin $P'_c$, possibly giving it a strong phase of
$O(1)$. Note that $|P'_c/{\widetilde T}'| \lsim 10\%$. That is, in the
SM the diagram responsible for the rescattering is considerably larger
than the diagram which receives the strong phase. On the other hand,
the NP rescattering can only come from the NP matrix elements
themselves. Assuming the same suppression factor, the NP strong phases
are $O(10\%)$, which is negligible, to a good approximation. Note that
this is a quite general result and applies to all NP models.

The neglect of NP strong phases allows for a great simplification.
For a given type of transition, all NP matrix elements can now be
combined into a single NP amplitude, with a single weak phase:
\beq
\sum \bra{f} {\cal O}_{\sss NP}^{ij,q} \ket{B} = \ANPq e^{i \Phi_q} ~,
\eeq
where $q=u,d,s,c$. (Throughout the paper, the symbols ${\cal A}$ and
$\Phi$ denote the NP amplitudes and weak phases, respectively.)
$\btopik$ decays involve only NP parameters related to the quarks $u$
and $d$. These operators come in two classes, differing in their
colour structure: ${\bar s}_\alpha \Gamma_i b_\alpha \, {\bar q}_\beta
\Gamma_j q_\beta$ and ${\bar s}_\alpha \Gamma_i b_\beta \, {\bar
q}_\beta \Gamma_j q_\alpha$ ($q=u,d$). The matrix elements of these
operators can be combined into single NP amplitudes, denoted
$\ApNPqph$ and $\ApNPCqph$, respectively \cite{BNPmethods}. Here,
$\Phi'_q$ and $\Phi_q^{\prime {\sss C}}$ are the NP weak phases; the
strong phases are zero. Each of these contributes differently to the
various $\btopik$ decays. In general, ${\cal A}^{\prime,q} \ne {\cal
A}^{\prime {\sss C}, q}$ and $\Phi'_q \ne \Phi_q^{\prime {\sss
C}}$. Note that, despite the ``colour-suppressed'' index $C$, the
matrix elements $\ApNPCqph$ are not necessarily smaller than the
$\ApNPqph$.

The $\btopik$ amplitudes can now be written in terms of the SM
amplitudes to $O({\bar\lambda})$ [$\pewp$ and $T'$ are related as in
Eq.~(\ref{pewprel})], along with the NP matrix elements
\cite{BNPmethods}:
\bea
\label{BpiKNPamps}
A^{+0} &\!\!=\!\!& -P' + \ApNPCdph ~, \nn\\
\sqrt{2} A^{0+} &\!\!=\!\!& P' - T' \, e^{i\gamma} - \pewp \nn\\
& & \hskip15truemm +~\ApNPcomb - \ApNPCuph ~, \nn\\
A^{-+} &\!\!=\!\!& P' - T' \, e^{i\gamma} - \ApNPCuph ~, \nn\\
\sqrt{2} A^{00} &\!\!=\!\!& -P' - \pewp + \ApNPcomb +
\ApNPCdph ~, 
\eea
where $\ApNPcomb \equiv - \ApNPuph + \ApNPdph$. There are now a total
of 11 theoretical parameters: $|P'|$, $|T'|$, $|{\cal A}^{\prime,
comb}|$, $|{\cal A}^{\prime {\sss C}, u}|$, $|{\cal A}^{\prime {\sss
C}, d}|$, $\gamma$, 3 NP weak phases and two relative strong
phases. With only 9 experimental measurements, it is not possible to
perform a fit. It is necessary to make some theoretical assumptions.

We assume that a single NP amplitude dominates. There are an infinite
number of choices, but we consider the following four possibilities:
(i) only ${\cal A}^{\prime, comb} \ne 0$, (ii) only ${\cal A}^{\prime
{\sss C}, u} \ne 0$, (iii) only ${\cal A}^{\prime {\sss C}, d} \ne 0$,
(iv) $\ApNPCuph = \ApNPCdph$, ${\cal A}^{\prime, comb} = 0$
(isospin-conserving NP).

In the first three cases there are seven parameters: three amplitude
magnitudes, $\gamma$, one weak NP phase and two relative strong
phases. However, for the type of NP characterizing the fourth fit, all
$\btopik$ amplitudes and their CP-conjugates contain two combinations
of amplitudes. These can be written as follows:
\bea
P_{\sss NP} e^{i\delta_{\sss NP}} e^{i\Phi_{\sss NP}} & = & -P'
+ \ApNPCdph ~, \nn\\
{\bar P}_{\sss NP} e^{i\delta_{\sss NP}} e^{-i\Phi_{\sss NP}} & = &
-P' + {\cal A}^{\prime {\sss C}, d} e^{-i
\Phi_d^{\prime C}} ~,
\eea
with $P_{\sss NP} \ne {\bar P}_{\sss NP}$. However, note that the real
parts of these quantities are equal: $P_{\sss NP} \cos(\delta_{\sss
NP} + \Phi_{\sss NP}) = {\bar P}_{\sss NP} \cos(\delta_{\sss NP} -
\Phi_{\sss NP})$. Thus, one variable, say $\delta_{\sss NP}$, can be
written as a function of the other three. That is, in this case there
is one fewer degree of freedom, and the $\btopik$ amplitudes contain
six unknown parameters: $P_{\sss NP}$, ${\bar P}_{\sss NP}$, $T'$,
$\gamma$, $\Phi_{\sss NP}$ and $\delta_{T'}$.

In cases (i), (iii) and (iv), there are more measurements than
unknowns, and we can perform a fit. On the other hand, parametrization
(ii) makes the same predictions as the SM: $A_{dir}(B^+ \to \pi^+K^0)
= A_{dir}(B^0\to \pi^0K^0) = 0$ and $A_{indir}(B^0\to \pi^0K^0) = \sin
2\beta$. Thus, in this case, there are only six observables, and we
cannot determine the seven theoretical parameters.

Using Table~\ref{bpiktable}, our results are: (i)
$\chi^2_{min}/d.o.f. = 1.9/2$ (39\%), (iii) $\chi^2_{min}/d.o.f. =
9.4/2$ (0.9\%), (iv) $\chi^2_{min}/d.o.f. = 3.9/3$ (27\%). Thus, based
on the fit quality only, we conclude that fit (i) is acceptable, fit
(iv) is marginal, and fit (iii) is poor.

However, these fits also give values for the CP angle $\gamma$: (i)
$\gamma = 64.2^\circ$, (iii) $\gamma = 31.8^\circ$, (iv) $\gamma =
37.8^\circ$. These can be compared with the value obtained from a fit
to independent data, $\gamma = {62^{+10}_{-12}}^\circ$
\cite{CKMfitter}. Note that this latter value of $\gamma$ includes
limits on $\bs$--$\bsbar$ mixing. However, the NP considered here
will, in general, also lead to effects in this mixing. Thus,
technically, in considering this type of NP, the $\bs$--$\bsbar$
mixing data should be removed from the $\gamma$ fit. In practice,
though, this will not make much difference. We therefore continue to
use the best-fit values of $\gamma$ with $\gamma =
{62^{+10}_{-12}}^\circ$ as the independent value, but the reader
should keep this caveat in mind.

Note also that any explanation of the $\btopik$ data using new physics
must also reproduce the SM value of $\gamma$. This demonstrates that,
in looking for NP, it is important to use all handles available, and
not simply concentrate on measurements of the CP phases.

We incorporate the information on $\gamma$ by adding a constraint to
the data, so that we now fit to all the $\btopik$ data and $\gamma =
{62 \pm 11}^\circ$. (Note that the $\gamma$ constraint is not a true
experimental number -- it has some theoretical input -- and so its
inclusion in the fit must be viewed with some prudence.) With this
added input, we can now perform a fit in parametrization (ii). We
find: (i) $\chi^2_{min}/d.o.f. = 1.9/3$ (59\%), (ii)
$\chi^2_{min}/d.o.f. = 2.7/3$ (44\%), (iii) $\chi^2_{min}/d.o.f. =
9.4/3$ (2\%), (iv) $\chi^2_{min}/d.o.f. = 6.7/4$ (15\%). We conclude
that fits (i) and (ii) are good, while fit (iv) is poor, and fit (iii)
is ruled out. We do not consider fit (iii) further.

However, we have still not included all the information at our
disposal. In the fits, we find that (i) $\delta_{T'} = -58.4^\circ$,
(ii) $\delta_{T'} = -26.2^\circ$ or $68.8^\circ$, and (iv)
$\delta_{T'} = -47^\circ$. On the other hand, the diagram $T'$ is
governed by the CKM matrix elements $V_{ub}^* V_{us}$, and so its
strong phase can arise only from self-rescattering. Thus, like the
new-physics amplitudes, the strong phase of $T'$, $\delta_{T'}$, is
expected to be very small. This requirement gives us an additional
handle. We incorporate this by adding a constraint to the data: we
require $\delta_{T'} = {0 \pm 10}^\circ$. Since this is purely
theoretical, it is obviously not on the same footing as the $\btopik$
data. However, since we only want to see if particular NP models give
a good fit, it is sensible to include this information among the
constraints.

Including the constraint on $\delta_{T'}$, we find (i)
$\chi^2_{min}/d.o.f. = 2.2/4$ (70\%), (ii) $\chi^2_{min}/d.o.f. =
5.9/4$ (21\%), and (iv) $\chi^2_{min}/d.o.f. = 14.3/5$ (1\%). We
therefore find that (i) is a good fit, (ii) is disfavored, and (iv) is
a very poor fit.

Of the four new-physics models examined in this paper, only one
produces a good fit to the $\btopik$ data and the various imposed
constraints on $\gamma$ and $\delta_{T'}$. It is case (i), ${\cal
A}^{\prime, comb} \ne 0$. In this model, the best-fit values of the
theoretical parameters are $|T'/P'| = 0.22$ (in line with theoretical
expectations), $|{\cal A}^{\prime, comb}/P'| = 0.36$, $\Phi' =
100^\circ$, $\delta_{P'} = -10^\circ$. We therefore find that the NP
amplitude must be sizeable, with a large weak phase.

This class of NP models essentially corresponds to a modification of
the SM electroweak penguin amplitude, as explored in
Refs.~\cite{BFRS,MY,BCLL}. In Ref.~\cite{BFRS} the weak phase of the
electroweak penguin was modified, meaning that the NP operator is of
the form $(V-A)\times (V-A)$. Here, we allow any form for the
operator, so that this is a more general solution. NP models which can
lead to this include $Z$- and $Z'$-mediated flavour-changing neutral
currents \cite{BCLL,ZFCNC} or supersymmetry with $R$-parity breaking.

Fit (ii) (${\cal A}^{\prime {\sss C}, u} \ne 0$) is marginal, but not
completely ruled out (it does give a value for $|T'/P'|$ which is
about three times larger than expectations). This is a NP solution
which has not been considered before. It can arise, for example, in
supersymmetric models with $R$-parity breaking. Fit (iii) (${\cal
A}^{\prime {\sss C}, d} \ne 0$) yields a poor fit, and this class of
NP models can therefore be ruled out. We also rule out
isospin-conserving models of NP [fit (iv)]. These include new physics
whose principal effect is to generate an anomalous gluonic quadrupole
moment \cite{chromo}.

A word of caution: one has to be careful about ruling out particular
models of new physics. Any specific NP model will, in general, lead to
more than one NP operator, and the more general case can be used to
explain the $\btopik$ data.

To summarize, we have presented a study of the current $\btopik$ data.
The standard model (SM) has great difficulty accounting for these
measurements. Depending on the parametrization, one obtains a poor
fit, or values for the SM parameters which are greatly at odds with
our present understanding. For models of new physics (NP), we adopt a
model-independent, effective-lagrangian parametrization of the NP
effects. There are three possible (complex) NP parameters which can
affect $\btopik$ decays, denoted ${\cal A}^{\prime, comb}$, ${\cal
A}^{\prime {\sss C}, u}$ and ${\cal A}^{\prime {\sss C}, d}$. We
consider four classes of NP models: (i) only ${\cal A}^{\prime, comb}
\ne 0$, (ii) only ${\cal A}^{\prime {\sss C}, u} \ne 0$, (iii) only
${\cal A}^{\prime {\sss C}, d} \ne 0$, (iv) isospin-conserving NP:
$\ApNPCuph = \ApNPCdph$, ${\cal A}^{\prime, comb} = 0$. Of these, the
classes of models (ii), (iii) and (iv) are ruled out or disfavored.
Only model (i) explains the data satisfactorily. It corresponds to a
modification of the electroweak penguin (EWP) amplitude. Note that,
while other studies also consider specific models of NP in the EWP,
our analysis is completely model independent.

\vskip2truemm 
D.S. thanks D.L. for the hospitality of the Universit\'e de
Montr\'eal, where part of this work was done. The work of S.B., P.H.,
D.L. and A.D. was financially supported by NSERC of Canada and les
Fonds FQRNT du Qu\'ebec. The work of D.S. was supported by the
U.S. Department of Energy under grant No.\ DE-AC02-98CH10886.

\end{document}